\documentclass[fleqn,usenatbib]{mnras}

\usepackage{float}
\usepackage{txfonts}

\usepackage[T1]{fontenc}
\usepackage{ae,aecompl}

\usepackage{natbib}
\usepackage{graphicx}	

\newcommand{\methanol}{$\mathrm{CH_3OH}$}
\newcommand{\htwo}{$\mathrm{H_2}$}
\newcommand{\nhtwo}{$\mathrm{n_{H_2}}$}
\newcommand{\tk}{$\mathrm{T_k}$}
\newcommand{\td}{$\mathrm{T_d}$}

\newcommand{\jone}{J\!=\!1\!-\!0}
\newcommand{\jtwo}{J\!=\!2\!-\!1}
\newcommand{\jthree}{J\!=\!3\!-\!2}

\newcommand{\vzero}{\varv\!=\!0}
\newcommand{\vone}{\varv\!=\!1}

\title[CS($\vzero$) masers in W51 e2e]{On the pumping of the CS($\vzero$) masers in W51 e2e}

\author[van der Walt et al.]{
D.J. van der Walt$^{1}$,\thanks{E-mail: johan.vanderwalt@nuwu.ac.za}
A. Ginsburg$^{2}$,
C. Goddi$^{3}$
\\
$^{1}$Centre for Space Research,North West University, Potchefstroom, North West, South Africa\\
$^{2}$Department of Astronomy, University of Florida, P.O. Box 112055, USA\\
$^{3}$Department of Astrophysics/IMAPP, Radboud University Nijmegen, P.O. Box 9010, 6500 GL Nijmegen, The Netherlands
}

\date{Accepted XXX. Received YYY; in original form ZZZ}

\pubyear{2020}

\begin{document}
\label{firstpage}
\pagerange{\pageref{firstpage}--\pageref{lastpage}}
\maketitle

\begin{abstract}
We present the results of numerically solving the rate equations for the first 31
rotational states of CS in the ground vibrational state to determine the conditions under
which the $\jone$, $\jtwo$ and $\jthree$ transitions are inverted to produce maser
emission. The essence of our results is that the CS($\vzero$) masers are collisionally
pumped and that, depending on the spectral energy distribution, dust emission can suppress
the masers. Apart from the $\jone$ and $\jtwo$ masers the calculations also show that the
$\jthree$ transition can be inverted to produce maser emission. It is found that beaming
is necessary to explain the observed brightness temperatures of the recently discovered CS
masers in W51 e2e. The model calculations suggest that a CS abundance of a few times
$10^{-5}$ and CS($\vzero$) column densities of the order $10^{16}\,\mathrm{cm^{-2}}$ are
required for these masers.  The rarity of the CS masers in high mass star forming
  regions might be the result of a required high CS abundance as well as due to
  attenuation of the maser emission inside as well as outside of the hot core.
  
\end{abstract}




\section{Introduction}

Molecular maser emission has always been a fascinating phenomenon associated with various
astrophysical environments such as eg. evolved stars, star forming regions and supernova
remnants. As for high-mass star forming regions, this has been particularly emphasized
over the last two decades through the discovery of periodic class II methanol,
formaldehyde and OH masers associated with high mass star forming regions \citep[see
  eg.][]{Goedhart2003, Goedhart2004, Goedhart2014, Goedhart2019, Araya2010, Szymczak2015, Szymczak2016,
  Szymczak2018, Sugiyama2017, Olech2019}. More recently water and methanol
maser flaring events associated with high mass star forming regions attracted significant
attention as such flaring events might be associated with episodic accretion outbursts
\citep{Brogan2018, Szymczak2018}. The periodic masers and maser flaring events are just
two examples which illustrate how masers can reveal changes in the star forming
environment which otherwise would not be detected with thermal line emission.

While widespread masers such as eg. $\mathrm{H_2O}$ and $\mathrm{CH_3OH}$ are useful in
studying some aspects of high mass star forming regions, rare masers, on the other hand,
must reveal something special about the star forming regions with which they are
associated. Amongst those masers for which until recently only a single detection has been
reported is that of CS. In the course of studying vibrationally excited emission from the
carbon-rich giant IRC +10216, \citet{Highberger2000} noted that the CS($\vone$) $\jthree$
transition has two components with exceptionally small line widths of respectively 1.1 and
2.5 $\mathrm{km\, s^{-1}}$ compared to other lines they detected. In contrast, the line
widths for other molecules reflect the full expansion velocity of 14.5
$\mathrm{km\,s^{-1}}$ of the gas in IRC+10216. In addition to the small line widths, the
brightness temperature of the CS($\vone$) $\jthree$ transition was found to be 526 K
compared to significantly smaller values for the other transitions of CS. This led
\citet{Highberger2000} to conclude that non-LTE conditions affect the CS ($\vone$)
$\jthree$ transition and that it might be a weak maser.

Recently, \citet{Ginsburg2019} reported the first discovery of maser emission for the
CS($\vzero$) $\jone$ and $\jtwo$ transitions associated with a high-mass young stellar
object, viz. W51 e2e. Although the observed line widths for the two masers of respectively
7.3 and 5.3 $\mathrm{km\,s^{-1}}$ were determined by the experimental velocity resolution,
both transitions have brightness temperatures of about 6700 K which rules out LTE
excitation. The two masers were found to be spatially separated by about 190 au and
spectrally by about 30 $\mathrm{km\,s^{-1}}$. \citet{Ginsburg2019} also imaged all three
hot cores in the W51 A region in the CS($\vone$) $\jone$ transition with a sensitivity of 1.1
$\mathrm{mJy~beam^{-1}}$ but did not detect it. This suggests that the physical
environment of the CS($\vzero$) $\jone$ and $\jtwo$ masers in W51 e2e is such that
excitation to the $\vone$ state is either weak or completely absent.

This detection of CS maser emission associated with a very young high-mass star raises the
questions of what the pumping mechanism is and why these masers are so rare. The aim of
this paper is first to investigate the pumping mechanism and based on that consider why
these masers seems to be rare. We restrict our investigation here to the CS($\vzero$)
masers in W51 e2e, leaving the case of the $\vone$ masers, as in IRC +10216, for later
follow-up work.

\section{Numerical method and molecular data}

The level populations were found by solving the well known set of rate equations in the
escape probability approach given by
\begin{eqnarray}\nonumber
\frac{dN_i}{dt} & = & \sum_{j < i}[(-N_i + (\frac{g_i}{g_j}N_j -
  N_i)W\mathcal{N}_{ij})\beta_{ij}A_{ij} \\  
\nonumber & &  + C_{ij}(N_j\frac{g_i}{g_j}e^{-E_{ij}/{kT}} - N_i)]\\
\nonumber  & & +  \sum_{j > i}[(N_j + (N_j  -
\frac{g_j}{g_i}N_i)W\mathcal{N}_{ji})\beta_{ji}A_{ji} \\ 
& & + C_{ji}(N_j - N_i\frac{g_j}{g_i}e^{-E_{ji}/{kT}})]
\label{eq:rate}
\end{eqnarray}
where $N_i$ is the number density in level $i$, $g_i$ the statistical weight of level $i$,
$A_{ij}$ the Einstein-A coefficient for spontaneous emission between levels $i$ and $j$,
$W$ the geometric dilution factor for an external radiation field and $\mathcal{N}_{ij}$
the photon occupation number for this field at frequency $\nu_{ij}$. $C_{ij} = n_{H_2}
K_{ij}$ is the collision rate with $n_{H_2}$ the \htwo{} number density and $K_{ij}$ the
collision rate coefficient. $\beta_{ij}$ is the escape probability. The rate equations
were supplemented with the particle number conservation requirement
\begin{equation}
N_{tot} = Xn_{H_2} = \sum_i N_i
\end{equation}
with $X$ the abundance relative to \htwo{} of the molecule under consideration.

The maser region is assumed to be embedded in or surrounded by a layer of warm dust at
temperature \td{}.  We assume that the dust radiation field is given by $ I_{\nu} =
\tau_d(\nu)B_\nu(T_d)$ with $B_\nu(T_d)$ the Planck function and $\tau_d(\nu) =
\tau_d(\nu_0)(\nu/\nu_0)^{p}$ the dust optical depth where we have set
$\tau_d(\nu_0=10^{13}\,\mathrm{Hz}) = 1$. \citep[see
  eg.][]{Pavlakis1996,Sobolev1997,Cragg2002}. We note that for dust emission
\citet{Sobolev1997} and \citet{Cragg2002} used $I_\nu \propto [1 -
  \exp(-\tau_d)]B_\nu(T_d)$ which, in the optically thin case is equivalent to
$(\nu/\nu_0)^pB_\nu(T_d)$.

We solve the rate equations using a fourth-order Runge-Kutta method, which is more
accurate and stable than Heun's method which was used in \citet{Vanderwalt2014}. The
initial distribution of the level populations was a Boltzmann distribution with
temperature equal to the average of the dust and kinetic temperatures.  For a given
\nhtwo{} the calculation started with a small CS specific column density such that the
system is completely optically thin in all transitions. An equilibrium solution for this
initial specific column density was then found by letting the system evolve in time steps
of 0.5 seconds till the convergence condition $|N_i(t_{j+1})-N_i(t_j)|/N_i(t_j) < 10^{-6}$
is reached for all levels $i$. The specific column density is then increased by a small
amount with the equilibrium solution for the previous value of the specific column density
being used as the initial distribution. The process is repeated until the required
specific column density is reached.

For the escape probability we used the expression for the large velocity gradient
approximation, ie.

\begin{equation}
  \beta_{ij} = \frac{1 - e^{-\tau_{ij}}}{\tau_{ij}}
  \label{eq:escprob}
  \end{equation}
with the optical depth given by 
\begin{equation}
  \tau_{ij} =
\frac{A_{ij}}{8\pi}\left(\frac{c}{\nu}\right)^3\left(\frac{g_j}{g_i}x_i -
x_j\right)\frac{N_{col}}{\Delta \varv}.
\label{eq:tau}
\end{equation}
$x_i$ and $x_j$ are the fractional number densities of molecules in the lower level, $i$,
and upper level, $j$, respectively, $N_{col}$ the CS column density and $\Delta \varv$ the
line width. The ratio $N_{col}/\Delta \varv$ is the specific column density.

When a population inversion occurs, beaming becomes important in which case
Eq.\,\ref{eq:escprob} becomes
\begin{equation}
  \beta_{ij} = \frac{\Omega_{ij}}{4\pi}\frac{e^{\tau_{ij}} -1}{\tau_{ij}}
  \label{eq:elitzurbeam}
\end{equation}
\citep{Elitzur1992} with $\Omega_{ij}/4\pi$ the fractional beam solid angle.  According to
\citet{Elitzur1992}, $\Omega_{ij}/4\pi = (2 \tau_{ij})^{-1}$. This dependence suggests
that beaming only starts when $\tau_{ij} = 0.5$. On the other hand, \citet{Spaans1992}
argued that $\Omega_{ij}/4\pi \approx \tau_{ij}^{-2}$. We combined these two and assumed
the beaming factor to be of the form $\Omega_{ij}/4\pi = (2\tau_{ij} + 1)^{-\eta}$ to
allow for beaming to start as soon as a population inversion occurs as well as to consider
stronger beaming than that of \citet{Elitzur1992}.

\begin{figure}
  \centering \includegraphics[width=\columnwidth]{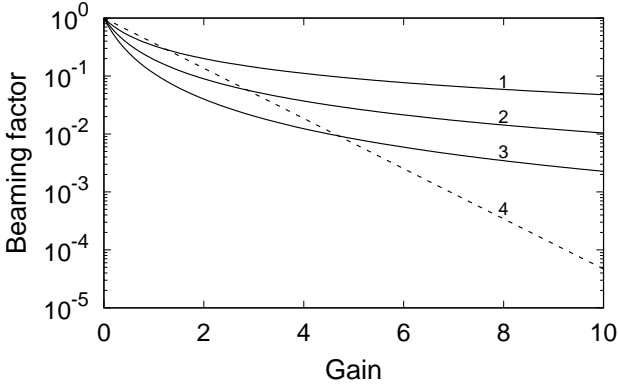}
  \caption{Comparison of the variation of the beaming factor for the four different
    dependencies of $\Omega/4\pi$ on the gain: (1) $(2\tau + 1)^{-1}$, (2)
    $(2\tau + 1)^{-1.5}$, (3) $(2\tau + 1)^{-2}$, (4) $10^{-0.433\tau}$ }
  \label{fig:taubeaming}
\end{figure}

We also considered the case when the beaming factor is that used by
\citet{Sobolev1997}. In their calculations, done within the framework of the large
velocity gradient approach, beaming was included through a parameter $\varepsilon =
\tau_\perp/\tau_0 = d(\ln\, \varv)/d(\ln\,r)$ with $\tau_0$ the optical depth in the
direction of the beam and $\tau_\perp$ the optical depth perpendicular to the direction of
the beam. The escape probability is then given by
\begin{equation}
  \beta(\tau, \varepsilon) = \int\limits_0^1\frac{1 + \varepsilon x^2}{\tau}\left[1 -
    \exp(-\frac{\tau}{1 + \varepsilon x^2})\right]dx
  \label{eq:sobolevbeam}
  \end{equation}
\citep{Castor1970}. $\beta(\tau, \varepsilon)$ can be evaluated as a function of $\tau$
for a given value of $\varepsilon$ and be equated to the right hand side of
Eq. \ref{eq:elitzurbeam} from which the corresponding value of $\Omega/4\pi$ can be found.
Following this procedure and taking $\varepsilon = 0.1$ \citep{Sobolev1997} it was found
that $\Omega/4\pi = 10^{-0.433\tau}$. In Fig.\,\ref{fig:taubeaming} we show the dependence
of the beaming factor on the gain ($\tau$) for $\eta = 1, 1.5, 2$ and the ``Sobolev'' beaming.

The Einstein-A and collisional rate coefficients of the first 31 levels of CS were taken
from the Leiden Atomic and Molecular Database \citep{Schoier2005}. The collisional rate
coefficients were extrapolated from 300 K to 500 K using equation 13 of \citet{Schoier2005} to
allow also for the possibility of higher gas kinetic temperatures  than what is given on
the Leiden Atomic and Molecular Database. 

The parameters involved in the present calculations are eg. \htwo{} density (\nhtwo), gas
kinetic temperature (\tk), dust temperature (\td), the geometric dust dilution factor
($W$), the index $p$ for the modified black body radiation and the beaming
factor. Exploring the entire parameter space (at least six dimensional) is the ideal but a
prohibitive task. However, information on parameters such as the geometric dilution, the
exact shape of the infrared radiation field and the beaming factor are basically not
available. To get a basic idea of the physical conditions under which a population
inversion can occur and how the maser brightness temperature vary, it turned out that
exploring parameter space consisting of \nhtwo{}, \tk{}, and \td{}, the values of which can be
estimated observationally, is sufficient. The maser brightness temperature is given by
\begin{equation}
  T_b = \frac{h\nu_{ij}}{k}\frac{x_i(1 - e^{-\tau_{ij}})}{(g_ix_j/g_j) - x_i}
  \label{eq:tb}
\end{equation}
with $i > j$ and $x_i$ and $x_j$ as in Eq.\,\ref{eq:tau}.

\section{Results}

\begin{figure}
  \centering \includegraphics[width=\columnwidth]{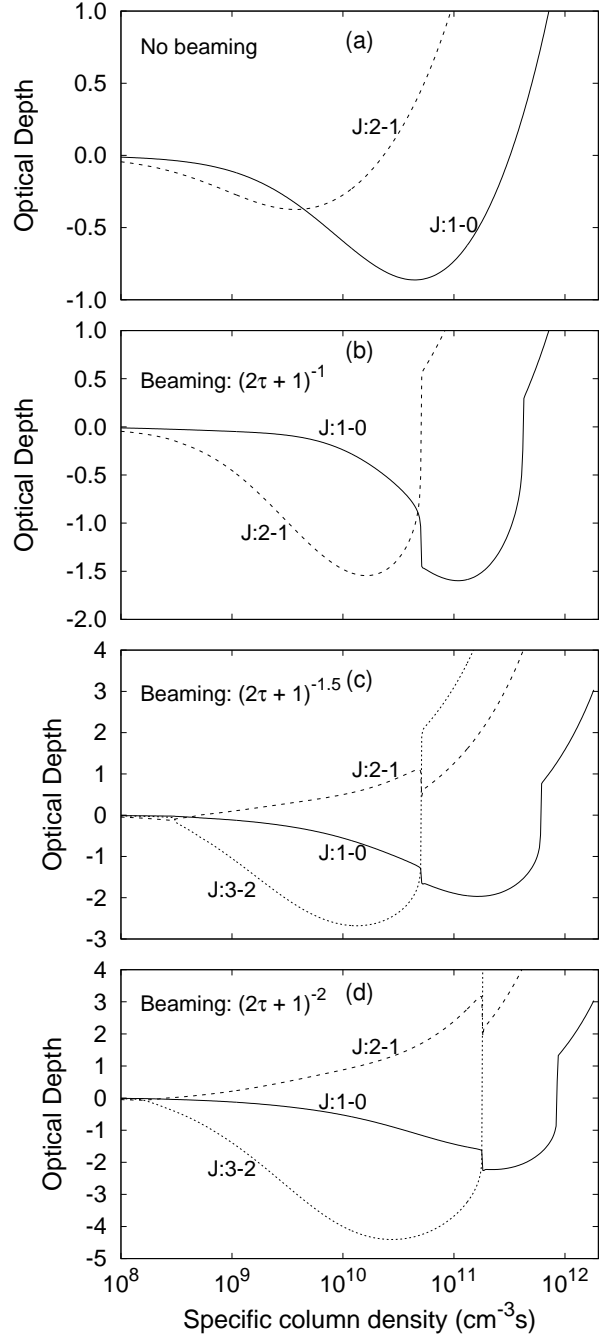}
  \caption{Figure to illustrate how the optical depth changes with specific column
    density as well as the effect of increased beaming. In all four panels only
    collisional excitation was considered. Note in (b) how the $\jone$ transition is affected
    when the inversion of the $\jtwo$ transition is lifted. Similar behaviour can also be seen
    in (c) and (d) where it is shown that the $\jthree$ transition is also inverted. Lifting of
    the inversion in the $\jthree$ transition affects both the $\jtwo$ and $\jone$
    transitions. Inversion of the $\jthree$ transition occurs only when collisional excitation
    is the dominant mechanism. Parameter values used: \nhtwo{} = $3.7\times
    10^{5}\,\mathrm{cm^{-3}}$, \tk{} = 300 K.    
  }
  \label{fig:tauvsscd02}
\end{figure}

As explained in the previous section, the calculational procedure to find the equilibrium
solution for a given combination of \nhtwo{}, \tk{}, \td{} and CS specific column density
is to start at a small specific column density where the system is optically thin, and to
evolve the system in small steps in specific column density. Fig.\,\ref{fig:tauvsscd02}
shows some examples of the variation of the optical depth as the specific column density
is increased in steps of 0.01 dex when excitation is through collisions and internal
generated radiation only, ie. the dust radiation field is switched off. All calculations
started at a specific column density of $10^6\,\mathrm{cm^{-3}\,s}$ and ended at $2\times
10^{12}\,\mathrm{cm^{-3}\,s}$.

Panel (a) shows the optical depth as a function of specific column density for the $\jone$
and $\jtwo$ transitions when there is no beaming. For both transitions the optical depth
decreases toward a minimum value (maximum gain) after which it increases to eventually
become positive. While both transitions can be inverted at the same specific column
density, the maximum gain for the $\jtwo$ transition is seen to occur at a smaller
specific column density than for the $\jone$ transition. As will be shown later it is
generally the case that the maximum gain for the two masers do not occur at the same
specific column density.

Panels (b) to (d) show how the behaviour of the optical depth changes when beaming is
introduced. Comparison with panel (a) shows that beaming has a marked effect on how the
optical depth changes with specific column density. Panel (b) shows that the specific
column density where maximum gain occurs has shifted to larger specific column
densities. Also, the maximum gain has increased. Perhaps the most interesting and
noteworthy difference with panel (a) is that the lifting of the inversion for the $\jtwo$
transition occurs over one step in specific column density. When this happens the gain of
the $\jone$ maser increases since population in the $J=1$ state ``suddenly''
increases. This behaviour is quite general since the two masers share the $J=1$ state in a
single ladder of rotational states.

A further interesting result of increasing the beaming is that the $\jthree$ transition
also becomes inverted as is seen in panels (c) and (d). Note how now three masers are
linked through the $J=1$, and $J=2$ states. Thus, when the inversion of the $\jthree$
transition is lifted, it affects the $\jtwo$ and the $\jone$ transitions. The inversion of
the $\jthree$ transition occurs only when collisions dominate the excitation and does not
occur at all combinations of \nhtwo{} and \tk{}. For most of what follows the focus will
be on the $\jone$ and $\jtwo$ masers in the ground vibrational state.

\begin{figure}
  \centering \includegraphics[width=\columnwidth]{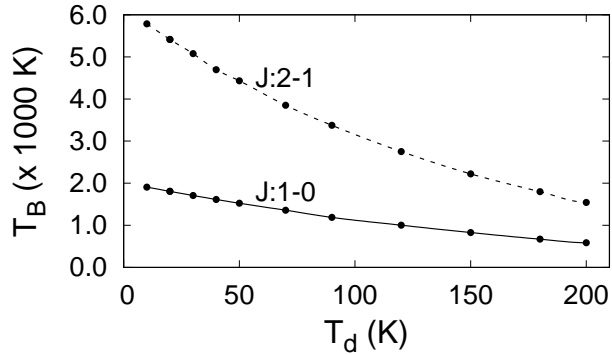}
  \caption{Graph showing how the brightness temperature changes with dust temperature for
    $\mathrm{n_{H_2} = 2.5\times 10^5\,\mathrm{cm^{-3}}}$, \tk{} = 360 K and a geometric dilution
    factor of 0.4. The emissivity of the dust was assumed to be that of a black body. The
    black dots indicate the dust temperatures at which the calculations were done. The
    solid and dashed lines are smooth fits to the points. The same applies to
    Fig.\,\ref{fig:dilution}. }
  \label{fig:td}
\end{figure}

\begin{figure}
  \centering \includegraphics[width=\columnwidth]{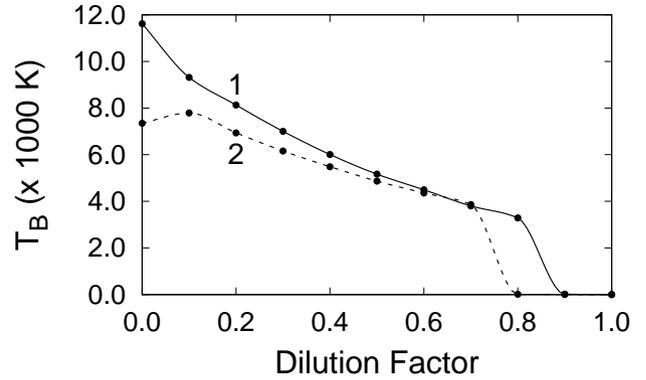}
  \caption{Graph showing how the brightness temperature changes with the geometric
    dilution of the dust radiation field. (1) For the $\jone$ transition when \nhtwo{} = $5
    \times 10^4\,\mathrm{cm^{-3}}$, \tk{} = 360 K, and \td{} = 100 K. (2) For the $\jtwo$
    transition when \nhtwo{} = $1.6 \times 10^5\,\mathrm{cm^{-3}}$, \tk{} = 460 K, and
    \td{} = 100 K. For both transitions the brightness temperatures are only a few Kelvin
    at the larger dilution factors.}
  \label{fig:dilution}
\end{figure}

The results presented in Fig.\,\ref{fig:tauvsscd02} show that inversion of the $\jone$ and
$\jtwo$ transitions can be accomplished with collisions as the primary excitation
mechanism which implies that the masers are collisionally pumped. An infra-red radiation
field, however, permeates the star formation environment and it is necessary to examine
what the effect of the presence of such a radiation field is. In addition to the temperature of
the dust, it is also necessary to consider the effect of geometric dilution. Instead of
considering the maser gain we will henceforth consider the maser brightness temperature,
calculated from Eq.\,\ref{eq:tb}, which is more relevant for comparison with
observations. The effect of the dust temperature on the brightness temperature is shown in
Fig.\,\ref{fig:td} and that of the geometric dilution in Fig.\,\ref{fig:dilution} when the
dust spectral energy distribution is that of a black body. Increasing the dust temperature
results in a decrease in the maser brightness temperature with a similar behaviour when
the dilution factor is increased. Note that a dilution factor of zero implies collisions
only and a value of one means that the dust emission originates inside the masing
region. These results imply that both masers are affected negatively when the dust
emission becomes stronger. The case when the dust emissivity is of the more general form
$(\nu/\nu_0)^p B_\nu(T_d)$ will be presented later.

A different representation which gives a better overall view of the behaviour of the two
masers can be obtained by considering the variation of the brightness temperature in the
\nhtwo{} - \tk{} plane. For this we considered \htwo{} densities in the range $10^4$ to
$10^6$ $\mathrm{cm^{-3}}$ with intervals of 0.1 dex. The kinetic temperature ranged
between 30 K and 500 K in steps of 10 K between 30 K and 300 K and steps of 20 K between
300 K and 500 K. This gives a total of 798 samplings in the \nhtwo{}-\tk{} plane. Unless
stated otherwise, a dust temperature of 100 K and a geometric dilution factor of 0.2 were
used. Presenting the model results in this way requires that a single brightness
temperature be associated with each combination of \nhtwo{} and \tk{}. In what follows it
is therefore assumed that the masers operate at maximum gain, ie. where the optical depth
has its largest negative value. The brightness temperature was calculated using the
corresponding level populations at maximum gain for the particular transition. The
brightness temperature was set equal to zero when there was no inversion.

\begin{figure}
  \centering \includegraphics[width=\columnwidth]{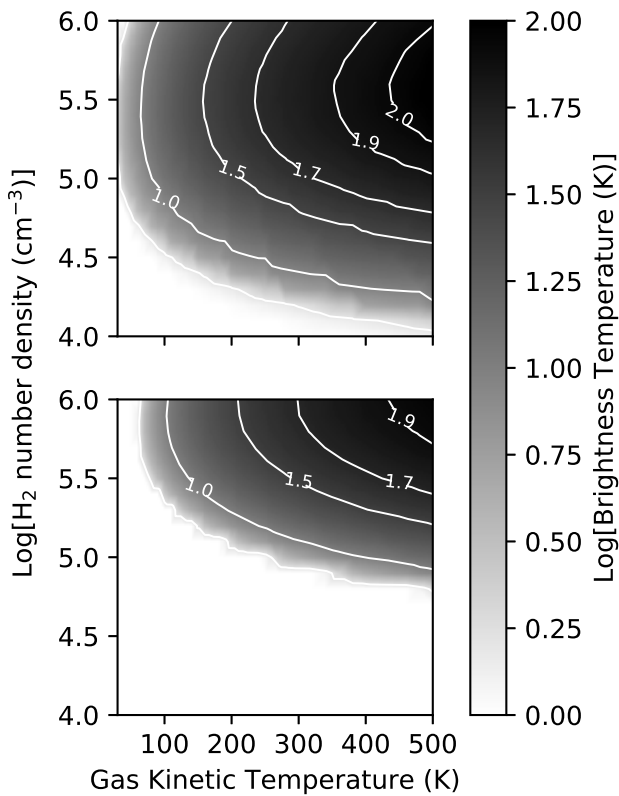}
  \caption{Contour and gray scale plot of the brightness temperature for the $\jone$
    transition (upper panel) and the $\jtwo$ transition (lower panel) in the \nhtwo{}-\tk{}
    plane when only collisions are included and with no beaming.  The contour levels in
    Figs.\,\ref{fig:nobeamcollonly} to \ref{fig:sobolevbeam} are in terms $\log
    \mathrm{T_B}$ }
  \label{fig:nobeamcollonly}
\end{figure}

In Fig. \ref{fig:nobeamcollonly} we show the variation of the brightness temperature when
collisions are the only excitation mechanism (infrared radiation field switched off) and
with no beaming.  It is seen that an inversion can obtained with collisions as the only
pumping mechanism over a significant part of the \nhtwo{}-\tk{} plane. Both panels suggest
that inversions occur also outside the ranges used for \nhtwo{} and \tk{}, ie. for
densities greater than $10^6 \mathrm{cm^{-3}}$ and $\mathrm{T_k} >$ 500 K. For the $\jone$
transition the maximum brightness temperature is 114 K and 91 K for the $\jtwo$
transition. The brightness temperatures for both transitions are significantly smaller
than what has been observed by \citet{Ginsburg2019}. Although the region over which there
is an inversion in the $\jtwo$ transition is completely overlapped by the region over
which there is inversion of the $\jone$ transition, it is again noted that the inversions
do not occur at the same CS specific column densities as will be discussed later.

\begin{figure}
  \centering \includegraphics[width=\columnwidth]{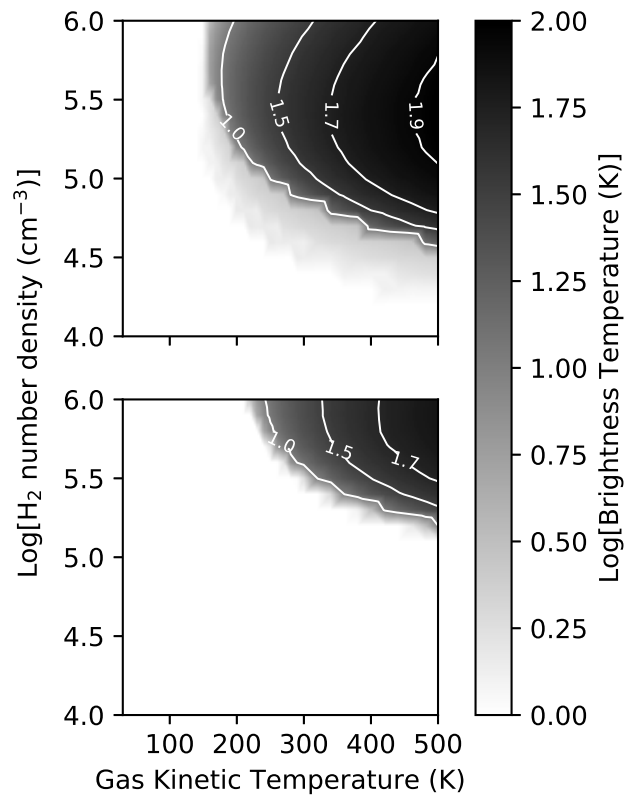}
        \caption{Contour and gray scale plot of the brightness temperature for the $\jone$
          transition (upper panel) and the $\jtwo$ transition (lower panel) in the
          \nhtwo{}-\tk{} plane in the presence of a black body dust radiation field with
          $\mathrm{T_d = 100\,K}$ and with no beaming.}
        \label{fig:nobeamcoltd100}
\end{figure}

Although radiative excitation alone does not result in inversion for any of the two
transitions, the presence of a dust radiation field does have an effect on the inverted
level populations as was shown in Figs.\,\ref{fig:td} and \ref{fig:dilution}.
Fig.\,\ref{fig:nobeamcoltd100} shows the variation of the brightness temperature when the
radiation field is that of a 100 K black body modified according to the wavelength
dependent escape probability. Comparison with Fig.\,\ref{fig:nobeamcollonly} shows that in
the presence of such a radiation field the brightness temperature at a specific
(\nhtwo{},\tk{}) point is reduced for both transitions. Also, for both transitions, the
region in the \nhtwo{}-\tk{} plane where inversion occurs is smaller and is shifted to
higher \htwo{} densities.
\begin{figure}
	\centering \includegraphics[width=\columnwidth]{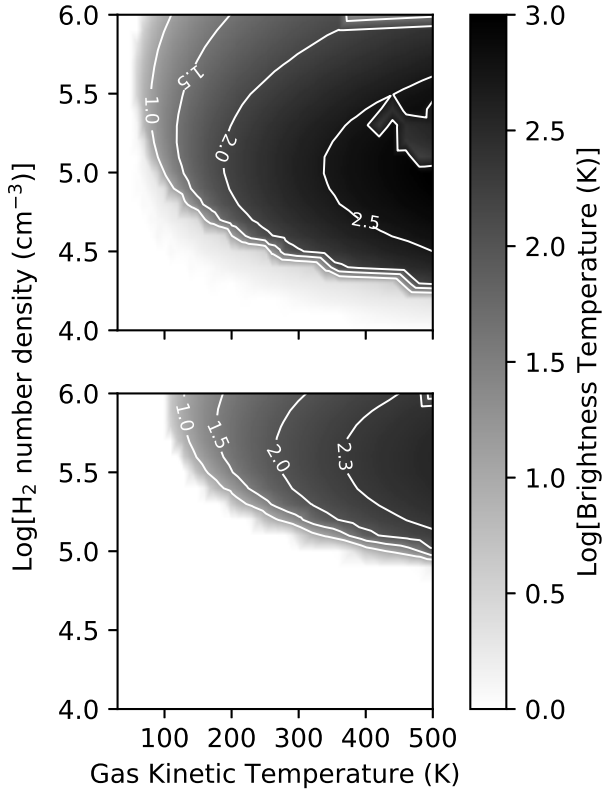}
        \caption{Contour and gray scale plot of the brightness temperature of the $\jone$
         transition (upper panel) and the $\jtwo$ transition (lower panel) in the \nhtwo
         -\tk$\,$ plane with beaming factor given by $(2\tau_{ij} + 1)^{-1}$ for $T_d$ =
         100 K and a geometric dilution factor of 0.2.}
    \label{fig:elitzbeam01}
\end{figure}

\begin{figure}
	\centering \includegraphics[width=\columnwidth]{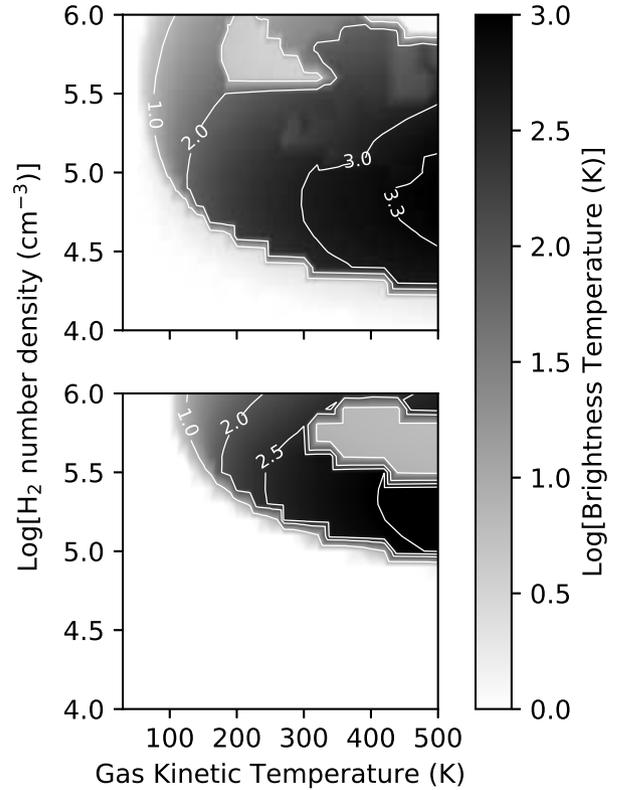}
        \caption{Same as for Fig.\,\ref{fig:elitzbeam01} but now 
         with the beaming factor given by $(2\tau_{ij} + 1)^{-1.5}$.}
    \label{fig:elitzbeam02}
\end{figure}

\begin{figure}
  \centering \includegraphics[width=\columnwidth]{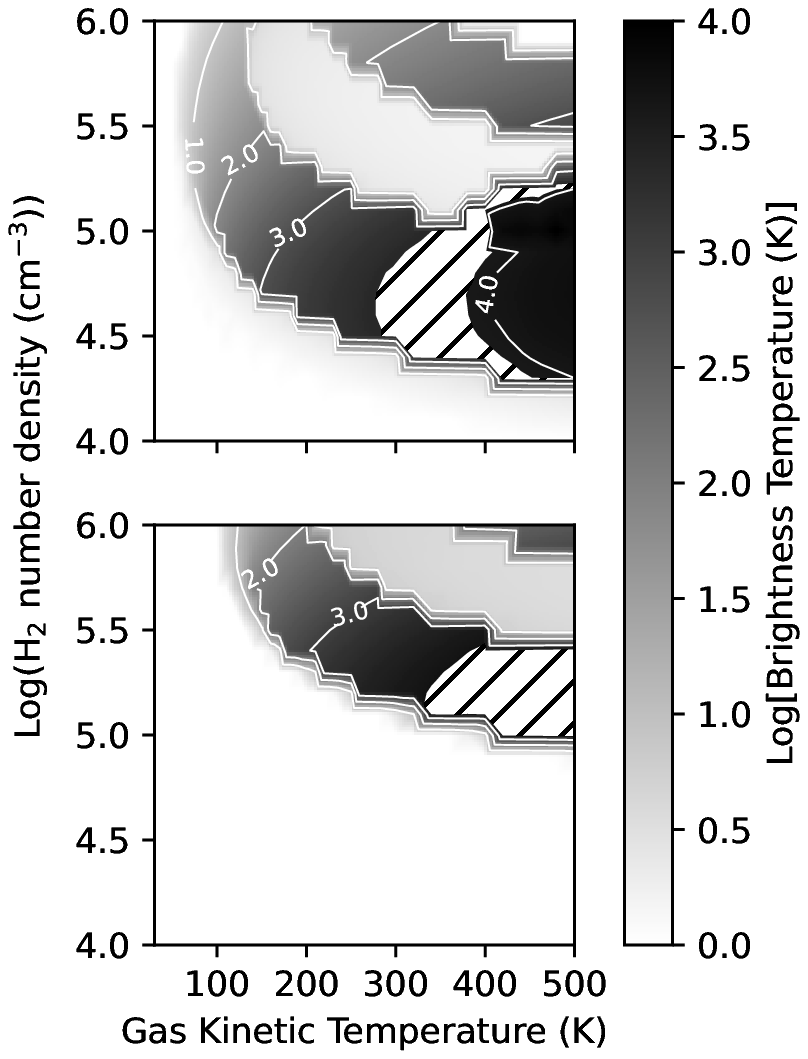}
        \caption{Same as for Fig.\,\ref{fig:elitzbeam02} but now with the beaming factor
          given by $(2\tau_{ij} + 1)^{-2}$.  The white filled region indicates the region
          where the maser brightness temperature lies between 6000 K and 8000 K.}
   \label{fig:elitzbeam03}
\end{figure}

In Figs.\,\ref{fig:elitzbeam01} to \ref{fig:elitzbeam03} we show the effect of beaming on
the maser brightness temperature when $\Omega_{ij}/4\pi = (2\tau_{ij} + 1)^{-\eta}$.
Comparison of Fig.\,\ref{fig:elitzbeam01} ($\eta=1$) with Fig.\,\ref{fig:nobeamcoltd100}
shows that not only is the region over which inversion occurs now larger but the
brightness temperatures are also higher. However, the maximum brightness temperatures of
593 K and 346 K for the $\jone$ and $\jtwo$ masers respectively, are still significantly
less than what has been observed for W51 e2e.  Increasing the beaming further
(Fig.\,\ref{fig:elitzbeam02}, $\eta=1.5$) results in significant changes in the shape of
the region in the \nhtwo{}-\tk{} plane where inversion occurs. In this case the maximum
brightness temperatures for the $\jone$ and $\jtwo$ masers are respectively 2550 K and
1400 K which are still too low to explain the observed brightness temperatures of the
masers in W51 2e2.

The result when $\Omega_{ij}/4\pi = (2\tau_{ij} + 1)^{-2}$ is shown in
Fig.\,\ref{fig:elitzbeam03}. The maximum brightness temperatures for the $\jone$ and
$\jtwo$ masers are respectively 24700 K and 8580 K. The white hatched regions show where
the brightness temperature for each maser lies between 5000 K and 9000 K. For the $\jone$
transition the region has an arch-like shape and lies more or less between kinetic
temperatures 300 K and 400 K and $10^{4.5} \lesssim \mathrm{n_{H_2}} \lesssim
10^5\,\mathrm{cm^{-3}}$. For the $\jtwo$ transition the corresponding values are
$\mathrm{T_k} \gtrsim 350$K and $10^5 \lesssim \mathrm{n_{H_2}} \lesssim
10^{5.5}\,\mathrm{cm^{-3}} $.  Figure \ref{fig:elitzbeam03} shows that with stronger
beaming the regions in the \nhtwo{}-\tk{} plane where the one maser is significantly
brighter than the other become separated.  Closer comparison of the upper and lower panels
shows that the $\jtwo$ transition is not inverted over most of the region where the
$\jone$ transition is inverted. The $\jone$ transition, on the other hand, is inverted
over most of the region where the $\jtwo$ maser occur but has brightness temperature which
is significantly less than that of the $\jtwo$ maser. Only at the very high temperatures
is there a small region where there is overlap. It is also seen that there are two regions
where the $\jone$ maser can occur, one at \htwo{} densities smaller and one at densities
greater than where the $\jtwo$ maser occur.

\begin{figure}
	\centering \includegraphics[width=\columnwidth]{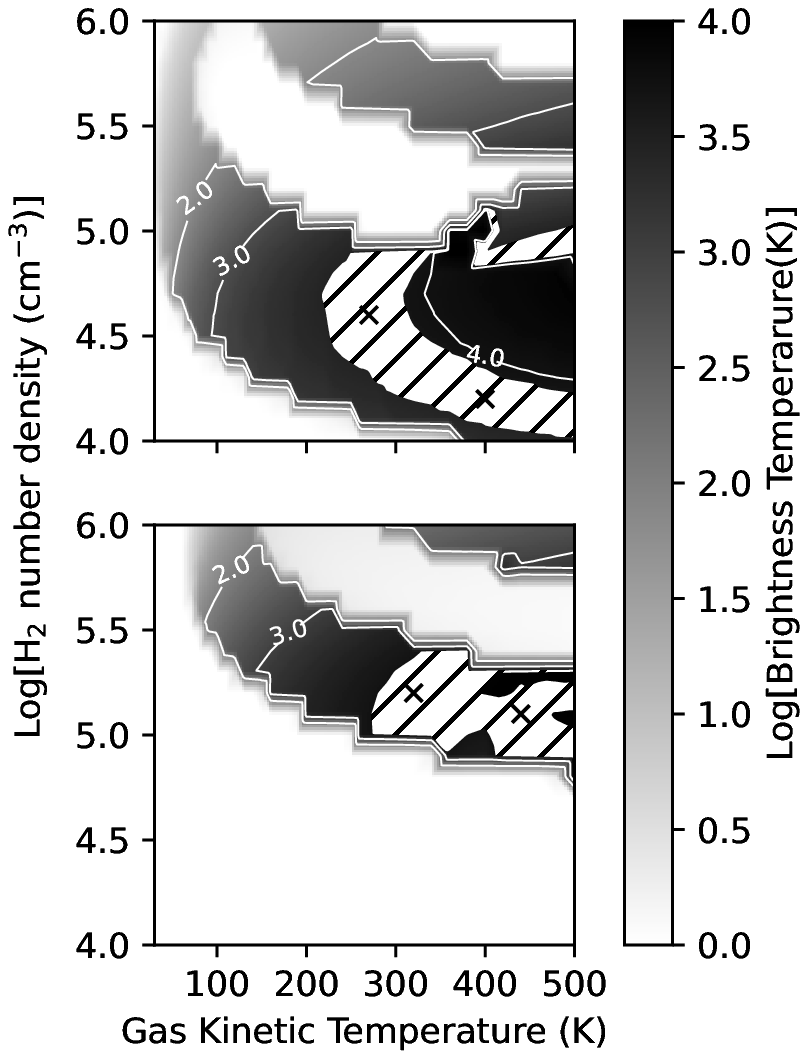}
       \caption{Same as for Fig.\,\ref{fig:elitzbeam03} but for a dust emissivity given by
         $(\nu/\nu_0)B_\nu(T_d)$. The hatched white region indicates the region where the
         maser brightness temperature lies between 6000 K and 8000 K. The four crosses
         indicate the combinations of \tk{} and \nhtwo{} used to estimate CS column
         densities and abundances.}
   \label{fig:beta1}
\end{figure}

\begin{figure}
	\centering \includegraphics[width=\columnwidth]{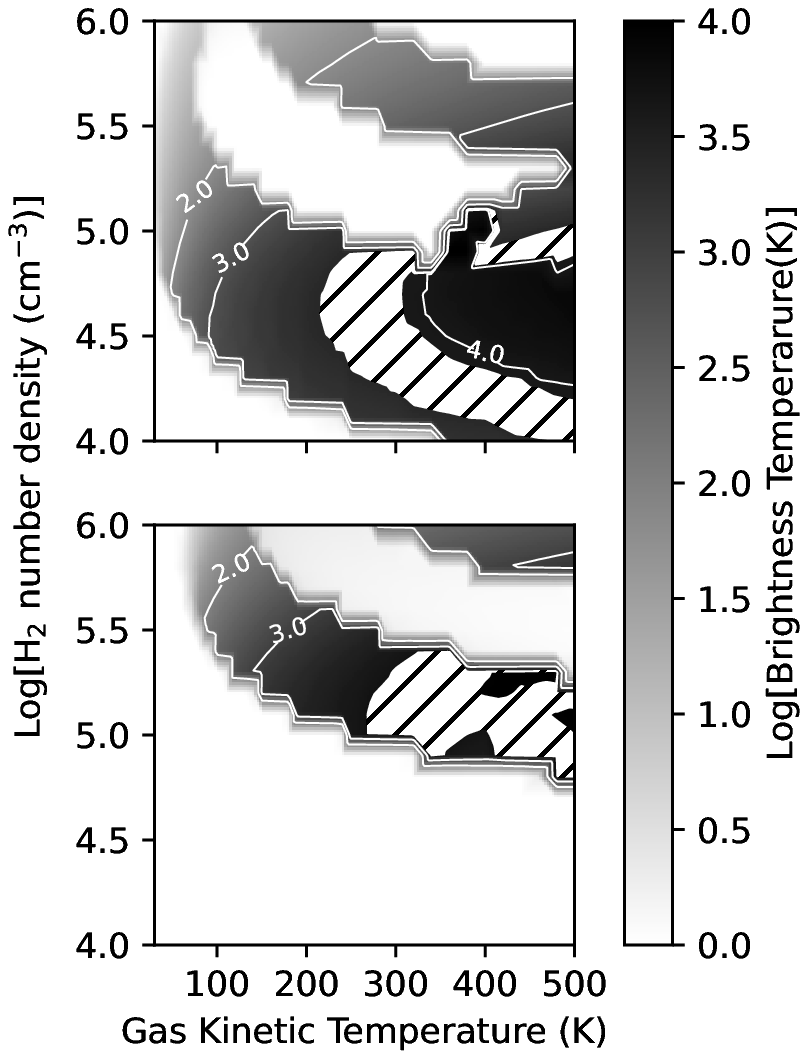}
       \caption{Same as for Fig.\,\ref{fig:elitzbeam03} but with no radiation field
         included. Excitation is therefore purely collisional.}
   \label{fig:beam2collonly}
\end{figure}

We also considered the case when the dust emissivity has the form $(\nu/\nu_0)B_\nu(T_d)$
with $\nu_0 = 10^{13}$ Hz. Fig.\,\ref{fig:beta1} shows the result for the same values of
the other parameters as for Fig.\,\ref{fig:elitzbeam03}. It is interesting to note that
the region over which the $\jone$ transition is inverted is significantly larger compared
to the case when it is assumed that the dust emissivity behaves like a black body. The
arch-like region where $5000 < \mathrm{T_B} < 9000$ for the $\jone$ transition in
Fig.\,\ref{fig:elitzbeam03}, now extends down to lower \htwo{} densities and higher
kinetic temperatures. To understand this change note that the largest energy difference
amongst the 31 levels ie. between $J=30$ and $J=29$, is equivalent to a frequency of about
$1.5 \times 10^{12}$ Hz. Having a dust emissivity that varies as $(\nu/\nu_0)B_\nu(T_d)$
with $\nu_0 = 10^{13}$ Hz means that the energy density in the radiation field at those
frequencies that can affect the level populations, is reduced by more than 15\% compared
to that of a black body emissivity. The effect of this change in the spectral energy
distribution is that collisions and spontaneous emission are effectively the only
mechanisms that affect the level populations. To confirm this we also performed
calculations when collisions are the only excitation mechanism and with the beaming factor
given by $(2\tau_{ij} + 1)^{-2}$. The result is shown in
Fig.\,\ref{fig:beam2collonly}. The similarity with Fig.\,\ref{fig:beta1} is obvious and
implies that dust emissivities of the form $(\nu/\nu_0)^p B_\nu(T_d)$ with $1 \le p \le
  2$, which is often used \citep[eg.][]{Pavlakis1996, Pavlakis2000, Cragg2002}, will
  result in the $\jone$ and $\jtwo$ masers having the same behaviour as that in
  Fig.\,\ref{fig:beam2collonly}.

\begin{figure}
	\centering \includegraphics[width=\columnwidth]{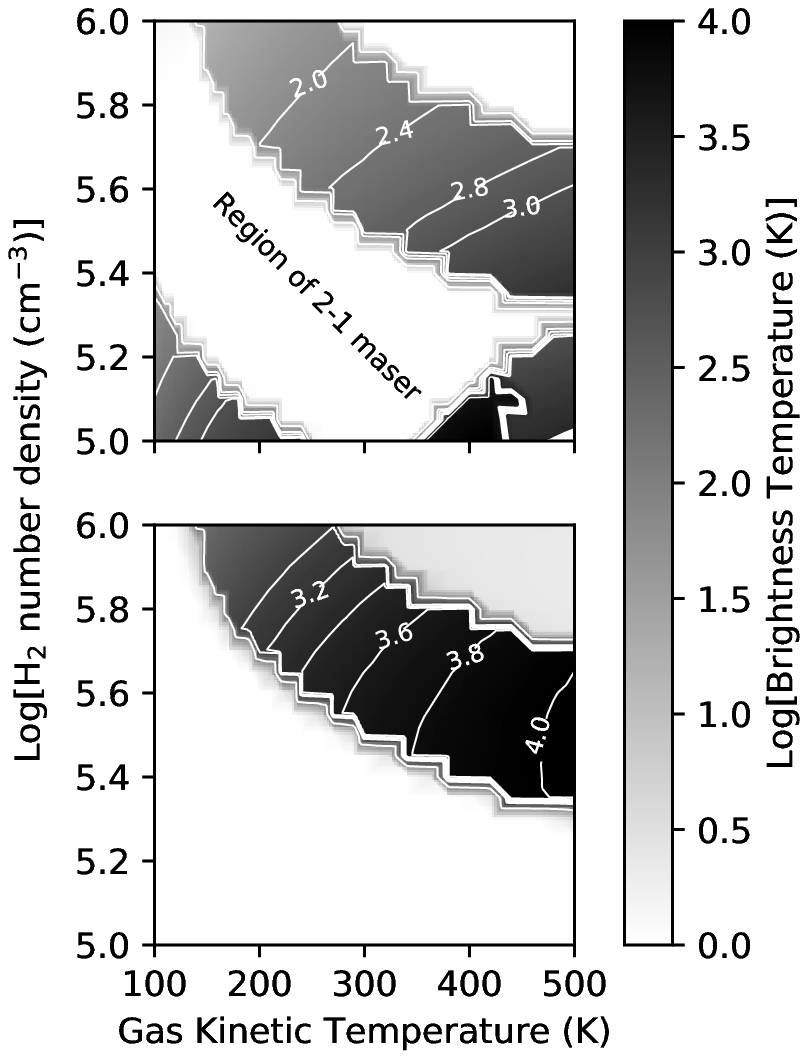}
        \caption{Contour and gray scale plot of the brightness temperature for the $\jone$
          transition (upper panel) and the $\jthree$ transition (lower panel) in the
          \nhtwo{}-\tk{} plane. The dust temperature was taken as 100 K with the
          emissivity given by $(\nu/\nu_0)^{1.5}B_\nu(T_d)$ where $\nu_0 = 10^{13}$. The
          beaming factor used was the same as in Fig.\,\ref{fig:elitzbeam03}.}
   \label{fig:tb43}
\end{figure}

A noticeable feature in Figs.\,\ref{fig:elitzbeam03} - \ref{fig:beam2collonly} is the
stair step behaviour which needs some explanation. Closer inspection of the model results
shows that at a given kinetic temperature, the brightness temperatures of the $\jone$ and
$\jtwo$ transitions increase significantly with an increase of 0.1 or 0.2 dex from the
density where there first is an inversion. For example, in the case of
Fig.\,\ref{fig:beam2collonly}, at \tk{} = 200 K the $\jone$ transition already has a very
weak inversion at \htwo{} densities of $10^4$ and $10^{4.1}\,\mathrm{cm^{-3}}$ with
brightness temperatures of 0.5 and 1.0 K respectively. Increasing the density to
$10^{4.2}\,\mathrm{cm^{-3}}$ raises the brightness temperature to 2833 K. The same
behaviour occurs at 180 K where the brightness temperature changes from 0.8 K at an
\htwo{} density of $10^{4.1}\,\mathrm{cm^{-3}}$ to 2215 K at $10^{4.2}\,\mathrm{cm^{-3}}$.
At 240 K the increase is from 1.3 K to 3300 K. Although there is an increase of about 1000
K in the brightness temperature from \tk{} = 180 K to 240 K at \nhtwo{} =
$10^{4.2}\,\mathrm{cm^{-3}}$, such a change is not visible on the contour plots. A contour
level of, say, 2000 K for the brightness temperature will therefore occur at the same
\htwo{} density for kinetic temperatures between 180 K and 240 K in this particular
example. The same explanation applies to other kinetic temperature intervals, eg. from 250
K to 340 K, where the maser brightness temperature, as represented by the contours, seems
to stay constant. The stair step effect is therefore a consequence of the 0.1 dex step
size in \nhtwo{} and would have been smaller if a smaller step size was used.  We note that
such rapid increase in brightness temperature with a small change in \nhtwo{} is also seen
in the results of \citet{Pavlakis2000} on OH masers.

\begin{figure}
	\centering \includegraphics[width=\columnwidth]{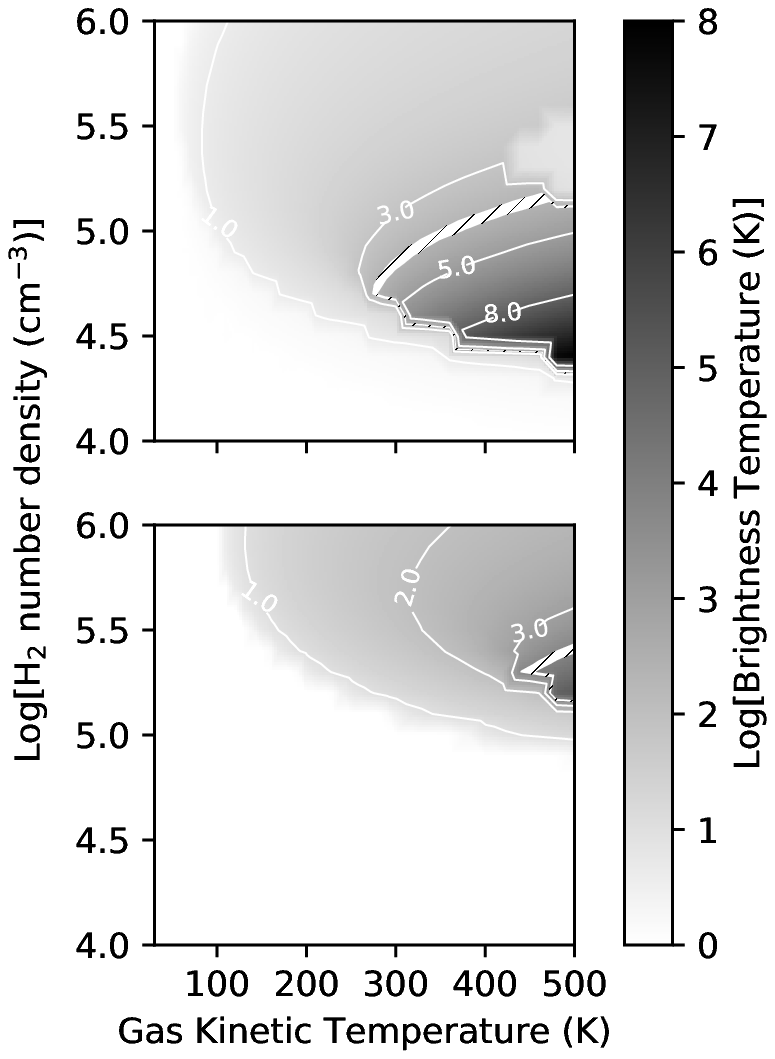}
       \caption{Contour and gray scale plot of the brightness temperature for the $\jone$
         transition (upper panel) and the $\jtwo$ transition (lower panel) in the \nhtwo
         -\tk$\,$ plane with beaming factor given by $10^{-0.433\tau_{ij}}$ for $T_d$ =
         100 K, and geometric dilution factor of 0.2.}
   \label{fig:sobolevbeam}
\end{figure}

Considering the results presented in panels (c) and (d) of Fig.\,\ref{fig:tauvsscd02} as
well as that in Figs.\,\ref{fig:beta1} and \ref{fig:beam2collonly}, we examined over what
region in the \nhtwo{}-\tk{} plane the $\jthree$ transition is inverted. The result is
shown in Fig.\,\ref{fig:tb43}. Since the inversion occurs only at higher densities the
calculations were done for \htwo{} densities greater than $10^5~\mathrm{cm^{-3}}$ only and
$\mathrm{T_k \geq 100~K}$. In this particular case the step size in density was 0.05
dex. Comparison of the two panels shows that the region where the $\jthree$ transition is
inverted is basically the same as the higher density region (there is also a region at
lower densities not shown in this Figure) where the $\jone$ transition is inverted but not
the $\jtwo$ transition. However, as was shown in Fig.\,\ref{fig:tauvsscd02}, the specific
column density where the $\jthree$ transition has maximum gain is smaller than where the
$\jone$ transition has maximum gain. The second point to note is that the brightness
temperature of the $\jthree$ maser is comparable with that of the $\jone$ and $\jtwo$
masers observed in W51 e2e, ie. in excess of several thousand Kelvin. Thus, at least
theoretically, these results suggest that, based on the brightness temperature, the
$\jthree$ maser should be detectable in W51 e2e depending on the presence of gas with
densities $\sim\!10^{5.4}~\mathrm{cm^{-3}}$ to $\sim\!10^{5.8}~\mathrm{cm^{-3}}$ and with
kinetic temperatures greater than about 250 K.

We also considered the case when beaming is taken to be the same as that used by
\citet{Sobolev1997}. The result is shown in Fig.\,\ref{fig:sobolevbeam}. While to a large
extent the distribution in the \nhtwo{}-\tk{} plane of where inversion occurs is the same
as for the other cases considered, it seen that the maximum brightness temperatures for
the $\jone$ and $\jtwo$ transitions are much higher. In the case of the $\jone$ transition it exceeds
$10^8$ K! For both the $\jone$ and $\jtwo$ transitions the regions where $5000 \le \mathrm{T_B}
\le 9000$ K are two narrow strips.

\begin{figure}
	\centering \includegraphics[width=\columnwidth]{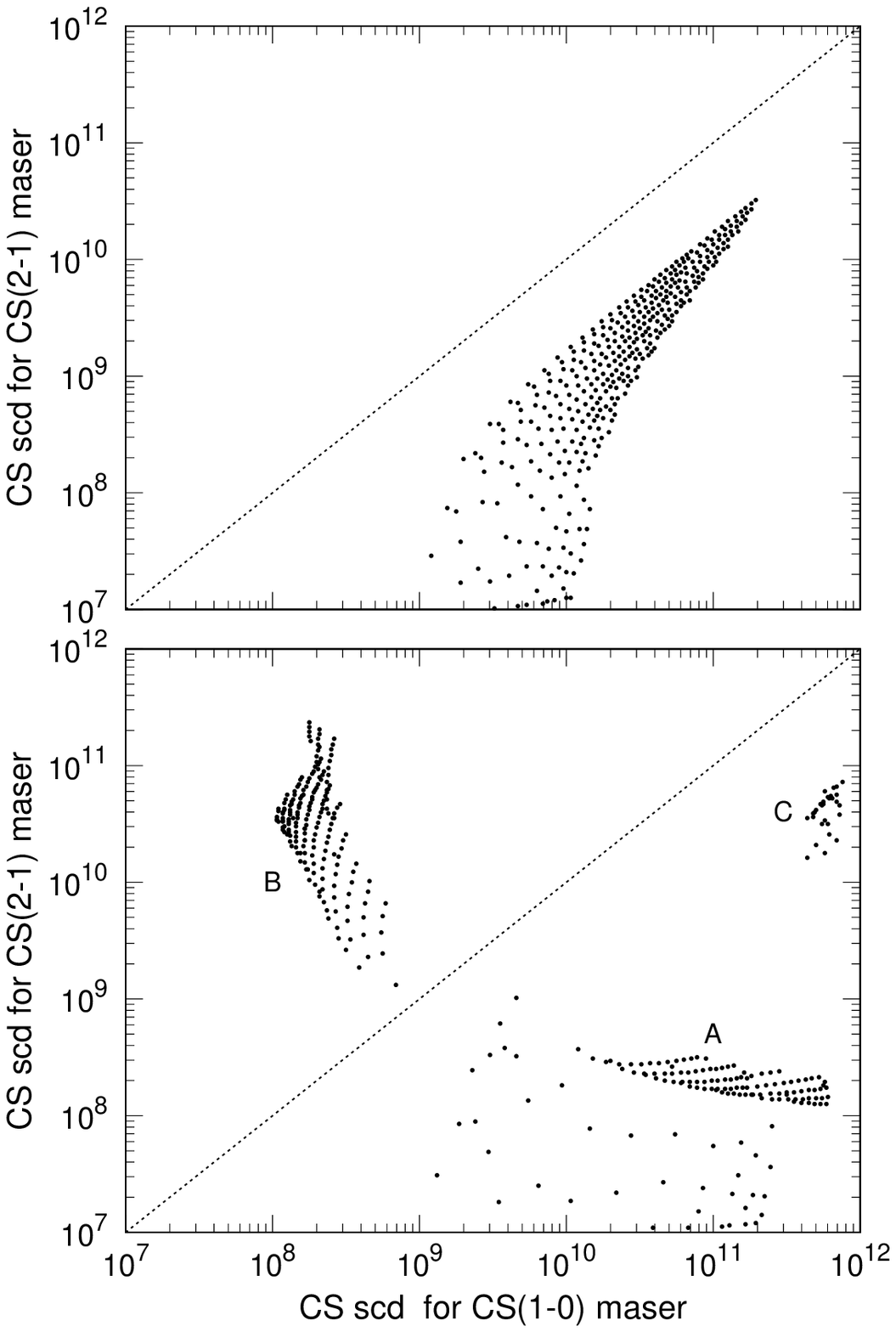}
        \caption{Plot of the specific column density where the $\jtwo$ maser has its
          maximum gain against the specific column density where the $\jone$ maser has its
        maximum gain. Upper panel: For collisions only and no beaming (see
        Fig.\,\ref{fig:nobeamcollonly}). Bottom panel: For collisions only and with the
        beaming factor given by $(2\tau_{ij} +1)^{-2}$ (see Fig.\,\ref{fig:beam2collonly}).}
   \label{fig:scds}
\end{figure}

It was already noted briefly with regard to Fig.\,\ref{fig:nobeamcollonly}, that although
there is overlap in the \nhtwo{}-\tk{} plane where the $\jone$ and $\jtwo$ masers occur,
these masers have their maximum gains at different specific column densities.  This is
illustrated in the upper panel of Fig.\,\ref{fig:scds} where the specific column density
where the $\jtwo$ maser has maximum gain is plotted against the specific column density
where the $\jone$ maser has maximum gain for those instances where \textit{both}
transitions are inverted.  All the points are seen to be located well below the $y=x$
line. Thus, for the case when collisions are the only excitation mechanism and with no
beaming, the specific column density where the $\jtwo$ maser has maximum gain is always
less than the specific column density where the $\jone$ maser as its maximum gain. In the
bottom panel of Fig.\,\ref{fig:scds} we show the same graph but for the case of
Fig.\,\ref{fig:beam2collonly} ie. with the beaming factor of the form $(2\tau_{ij} +
1)^{-2}$ and collisions as the only excitation mechanism. Refering to
Fig.\,\ref{fig:beam2collonly}, this is the case where there are rather well defined
regions in the \nhtwo{}-\tk{} plane where one of the masers is strong and the other one
weak or absent or both masers are strong. From the bottom panel of Fig.\,\ref{fig:scds} it
is seen that now there are three clearly distinct groups of points. Group A is for the
case when the maximum gain and the brightness temperature of the $\jone$ maser are
significantly greater than that of the $\jtwo$ maser. Typically, for group A the
brightness temperature of the $\jone$ maser is greater than 100 K while for the $\jtwo$
maser it is only about 1 or 2 K. Group B is just the opposite of group A, viz. the $\jtwo$
maser is bright and the $\jone$ maser very weak. For group C both masers have brightness
temperatures greater than 1000 K. The location of the three groups relative to the $y = x$
line shows that generally the $\jone$ and $\jtwo$ masers do not have their respective
maximum gains at the same CS specific column density.

\section{Discussion}

\subsection{General comments}
One of the first questions asked about a particular astronomical maser is how it is
pumped. In the case of the CS masers the results of the numerical calculations presented
above clearly suggest that these masers are pumped collisionally. It was shown that the
$\jone$, $\jtwo$ and $\jthree$ transitions can be inverted in the absence of an external
radiation field or when the dust emissivity is of the form $(\nu/\nu_0)^pB_\nu(T_d)$ with
$1 \le p \le 2$ and $\nu_0 = 10^{13}$ Hz, in which case the emission below $10^{13}$ Hz is
strongly reduced relative to that of a black body. On the other hand, dust emission with a
black body spectral energy distribution below about $10^{13}$ Hz negatively affects the
population inversion of these three transitions. Considering all of these results it can
be concluded that the CS masers are collisionally pumped.

We also note that there is some qualitative agreement with the observations of
\citet{Ginsburg2019}. These authors found that the $\jone$ and $\jtwo$ masers are clearly
separated spatially and kinematically. Considering the cases where the brightness
temperatures of the two masers fall between 5000 K and 9000 K
(Figs.\,\ref{fig:elitzbeam03} and \ref{fig:beta1}) as well as that the two masers don't
have maximum gain at the same specific column densities (Fig.\,\ref{fig:scds}) very
strongly suggests that the $\jone$ and $\jtwo$ masers will observationally not be found to
be co-spatial. 

As pointed out by \citet{Ginsburg2019}, the newly discovered CS $\jone$ and $\jtwo$ masers
associated with W51 e2e form part of a group of rare masers which raises the question of
why these masers are so rare. These authors speculate that W51 e2e may have unique
characteristics regarding the infrared radiation field, chemistry, geometry or any other
unique aspect that favours CS masers. As for the radiation field, we already noted that
brighter masers are produced in an environment with a dust emissivity of the form
$(\nu/\nu_0)^p B_\nu(T_d)$ rather than with a black body dust emissivity. However, such a
dust emissivity is not unique and cannot be considered as a reason why these masers are
rare.  Another factor may be the geometry of the masing region which can affect the
beaming and therefore the brightness of the masers. It was found that beaming which is
stronger than that given by \citet{Elitzur1992} but similar to that found by
\citet{Spaans1992} is needed to explain the observed brightness temperatures of the
masers. However it should not be as strong as that used by \citet{Sobolev1997}. On the
other hand, W51 e2e also has associated class II methanol masers which quite generally
have very high brightness temperatures and which requires strong beaming. Why a different
behaviour of the beaming factor is required for CS and $\mathrm{CH_3OH}$ masers is not
clear.

\subsection{Column densities}
Two other factors that can play a role in explaining the uniqueness of W51 e2e for
favouring CS masers are the CS abundance and column density. We first consider the column
density ($\mathrm{cm^{-2}}$) which is found from the CS specific column density where the
inversion is maximum and the linewidth. The linewidths given by \citet{Ginsburg2019} are
basically instrumental linewidths and it can therefore be expected that the real maser
linewidths are smaller. Here we use a linewidth of 1.0 $\mathrm{km\,s^{-1}}$ for both
masers which can be considered as a lower limit. For illustrative purposes, CS column
densities were calculated for both masers at two combinations of kinetic temperature and
\htwo{} density as indicated by the crosses in Fig.\,\ref{fig:beta1}. For the $\jone$
maser at \tk{} = 270 K and \nhtwo{} = $10^{4.6}\,\mathrm{cm^{-3}}$ the column density is
found to be $\log \mathrm{N_{CS}} = 15.9$. For \tk{} = 400 K and \nhtwo{} =
$10^{4.2}\,\mathrm{cm^{-3}}$ the column density is $\log \mathrm{N_{CS}} = 16.1$ which,
for practical purposes, is the same as for the lower temperature case. For the two
combinations for the $\jtwo$ maser the CS column density was found to be the same
viz. $\log \mathrm{N_{CS}} = 15.6$.

These values can be compared with column densities derived by a number of authors using LVG
model fitting of observations of high-mass star forming regions. For 11 northern cores
\citet{Zinchenko1994} found $\langle \log \mathrm{N_{CS}}\rangle =
13.6$. \citet{Juvela1996} did a multi transition study toward 33 southern $\mathrm{H_2O}$
maser sources and found $\langle \log \mathrm{N_{CS}}\rangle = 13.62$. \citet{Plume1997}
also did a multi transition CS study of 150 high-mass star forming regions with associated
$\mathrm{H_2O}$ masers. These authors found $\langle \log \mathrm{N_{CS}}\rangle = 14.36$
for 71 of these regions. \citet{Larionov1999} did a survey of 158 bipolar outflow and
methanol maser sources in the Northern sky and give CS column densities for 50 of these
sources with $\langle \log \mathrm{N_{CS}}\rangle = 14.47$. Simply taken at face value,
these column densities are at least an order of magnitude less than what is derived from
the model calculations. However, it has to be noted that all four of these studies were
using single dish telescopes and that the derived column densities are averages over
significantly large regions of the individual star forming regions observed. For example,
the beam size for the observations by \citet{Plume1997} range between 10 - 30 arcsec and
for W51 included many unresolved cores for which the CS column densities certainly are
higher than the beam averaged values.

\subsection{Abundances}
\label{sec:pathlengths}
As in the case for the column densities, a quantitative comparison of abundances derived
from observations with that from the model is not simple for two reasons. First,
abundances determined from single dish observations are, like column densities, averages
along the line of sight as well as beam averaged and therefore do not take source
structure into account \citep{Wakelam2004}. Second, abundances derived from the model
results are calculated from the relation $X_{\mathrm{CS}} \mathrm{n_{H_2}} \ell =
\mathrm{N_{CS}}$ where $\mathrm{X_{CS}}$ is the CS abundance relative to \htwo{}, $\ell$
the maser pathlength and, $\mathrm{N_{CS}}$ is the CS column density calculated from the
product of the specific column density where the maser gain is maximum and the maser
linewidth (taken as 1 $\mathrm{km\,s^{-1}}$).  The maser pathlength is unknown and the
derived CS abundance will therefore depend on the assumed maser pathlength. In spite of
these limitations, it is nevertheless useful to compare the abundances inferred from the
model calculations with some observationally derived values. We first consider CS
abundances derived from observations and then maser pathlengths used by other authors.

\citet{Dickens2000} derived CS abundance ranges between $7\times 10^{-10}$ and $7\times
10^{-9}$ for the dark cloud L134N. Such low abundance is expected for dark clouds. For the
Orion hot core \citet{Chandler1997} derived a CS abundance of $1.2 \times 10^{-8}$, which
is only slightly higher than the maximum CS abundance for L134N. Similar abundances were
also found for W43MM1, IRAS18264-1152, IRAS05358+3543, IRAS18162-2048 \citep{Herpin2009}
and AFGL2591 \citep{VanderTak2003}.  \citet{Bergin1997} derived CS abundances relative to
CO of between $2 \times 10^{-5}$ and $1 \times 10^{-4}$ for M17 and Cepheus respectively,
which gives CS abundances relative to \htwo{} of the order $10^{-8} - 10^{-9}$ using
$\mathrm{[CO]/[H_2]} = 10^{-4}$ \citep{Kruegel2003}. More recently
  \citet{Barr2018} derived a CS fractional abundance relative to \htwo{} of $2 \times
  10^{-6}$ for the hot molecular core associated with the massive protostar AFGL 2591
  using mid-infrared spectroscopy. These observations probed the hot gas ($\sim$ 700 K)
  which strongly points a higher CS abundance closer to the protostar.

As for maser pathlengths we note the following. Based on an angular diameter of 0.005
arcsec for OH masers in W3(OH), \citet{Moran1978} inferred a total maser pathlength of
about $10^{15}$ cm for unsaturated OH masers. \citet{Reid1980} similarly concluded that
gain lengths of the order of $2\times 10^{15}$ cm are required to explain very bright OH
masers in W3(OH). Also considering OH masers, \citet{Gray1992} concluded that long maser
path lengths, ie. greater than $10^{14}$ cm may be composed of a series of regions along
the line of sight having similar velocities. These authors also conclude that for OH
masers, pathlengths as short as $10^{13}$ cm may be adequate to explain observed maser
brightness temperatures. \citet{Pavlakis1996} and \citet{Pavlakis2000} used a maser
pathlength of $5 \times 10^{15}$ cm for OH masers in star forming regions. On the other
hand, \citet{Sobolev1997} and \citet{Cragg2002} assumed a pathlength of $10^{17}$ cm in
their models for class II \methanol{} masers.

In spite of the fact that deriving a CS abundance from the model results depends on an
assumed maser pathlength, some constraints can be used to put limits on the maser
pathlength and therefore on the CS abundance.  As already noted earlier, the $\jtwo$ maser
is pumped at a higher \htwo{} density compared to the $\jone$ maser which might suggest
that the $\jtwo$ masing region is closer to the exciting star and that the kinetic
temperature is also higher than in the masing region of the $\jone$ maser. To take such a
scenario approximately into account we calculated $\langle\mathrm{N_{CS}/n_{H_2}}\rangle$
over selected regions of the \nhtwo{}-\tk{} plane where the brightness temperature for the
$\jone$ and $\jtwo$ masers lie between 5000 K and 9000 K. For the $\jone$ maser we
considered \htwo{} densities greater than $10^{4.6}~\mathrm{cm^{-3}}$ and kinetic
temperature between 220 and 290 K. For the $\jtwo$ maser we assumed the kinetic
temperature to be between 300 and 400 K.  (see Figs.\,\ref{fig:beta1} and
\ref{fig:beam2collonly}).  Maser pathlengths can then be calculated using abundances based
on observations.

Thus, assuming a CS abundance of, say, $10^{-8}$, pathlengths for the $\jone$ and $\jtwo$
masers of respectively $1.5\times 10^6$ AU and $2.6 \times 10^5$ AU are found which,
obviously, are completely unrealistic. The implication is that the CS abundance associated
with the two masing regions must be greater than $10^{-8}$. Increasing the abundance to
$10^{-6}$ means that the corresponding maser pathlengths become $1.5 \times 10^4$ AU and
$2.6 \times 10^3$ AU respectively. A pathlength of $1.5 \times 10^4$ AU for the $\jone$
maser is also doubtful since it implies a maser pathlength significantly larger than the
outflows and the dust continuum emission associated with W51 e2e \citep[see Fig. 1
  of][]{Goddi2018}. The $\jone$ and $\jtwo$ masers are projected at the base of the
outflow and may be associated with a disk or other rotating structure. Given the size of a
possible disk as shown by \citet{Goddi2018}, it seems reasonable to argue that the longest
pathlength should be less than the diameter of the disk, ie. $<$ $\sim\,$720 AU. For
example, a CS abundance of $2.5 \times 10^{-5}$ results in pathlengths of $\sim$ 670 and
$\sim$ 120 AU for the $\jone$ and $\jtwo$ masers respectively. The pathlengths may even be
smaller due to radial velocity gradients in a rotating structure which means that the CS
abundance may also be higher. However, the CS abundance is expected to be less than the
general CO abundance of $\sim 10^{-4}$ which requires the pathlength for the $\jone$ maser
to be greater than about 150 AU and greater than about 26 AU for the $\jtwo$ maser. Even
though it is not possible to fix the maser pathlength, it is rather clear that the model
calculations require the CS abundance in W51 e2e to be at least two orders and maybe even
three orders of magnitude higher than the CS abundances based on single dish
observations. The higher CS abundance suggested by the model calculations is at least
qualitatively in agreement with the observational results of \citet{Barr2018}.

\subsection{Maser amplification or attenuation in the surrounding medium?}

The results in Figs. \ref{fig:elitzbeam03} - \ref{fig:beam2collonly} show the regions in
the \htwo{}-\tk{} plane where the brightness temperatures for $\jone$ and $\jtwo$ masers
corresponds to that observed by \citet{Ginsburg2019}. Considering the \htwo{} densities
and kinetic temperatures where the masers occur it follows that the maser emission has to
propagate through considerable molecular material before ``emerging'' from the star
forming region. The question therefore arises whether there is additional amplification or
attenuation of the masers in the surrounding molecular envelope.

\begin{figure}
	\centering \includegraphics[width=\columnwidth]{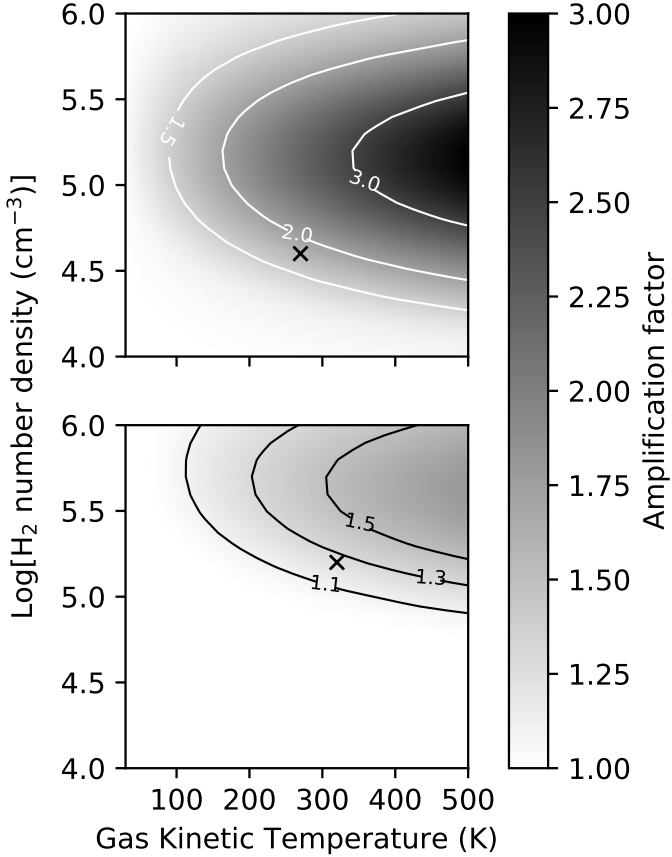}
        \caption{Contour and gray scale plot of the amplification factor ($e^ 
          {-\tau} $) for the $\jone$ transition (upper panel) and the $\jtwo$
          transition (lower panel) in the \nhtwo{}-\tk{} plane when only collisions are
          included and with no beaming.}
   \label{fig:amplify}
\end{figure}

Further amplification of the maser emission will take place if it propagates through
regions where a population inversion also exists. Our calculations do not take such a
scenario explicitly into account. Consider, however, for example the solutions at \tk{} =
270 K and \nhtwo{} = $10^{4.6}~\mathrm{cm^{-3}}$ for the $\jone$ maser and for the $\jtwo$
maser at \tk{} = 320 K and \nhtwo{} = $10^{5.2}~\mathrm{cm^{-3}}$ as indicated by the
crosses in the upper and lower panels of Fig.\,\ref{fig:beta1}. Quite generally it can be
assumed that maser emission produced under these conditions, when propagating radially
outward, will encounter a density and temperature gradient such that the density and
kinetic temperature decreases. If it is further assumed that no beaming occurs outside of
the masing region and that excitation is dominated by collisions, the situation as shown
in Fig.\,\ref{fig:nobeamcollonly} applies. Fig.\,\ref{fig:amplify} shows the variation of
the amplification factor ($e^{-\tau}$) corresponding to the brightness temperature as
shown in Fig.\,\ref{fig:nobeamcollonly}. It is seen that without any beaming further
amplification outside the masing region is rather small for both the $\jone$ and $\jtwo$
masers and that it decreases toward lower densities and kinetic temperatures.

It is also seen from Fig.\,\ref{fig:amplify} that there is a significant part in the
\htwo{}-\tk{} plane where, for both masers, there is no amplification which implies that
the maser beam might be attenuated when propagating through the surrounding molecular
medium. To estimate the attenuation we calculated (assuming LTE) the optical depth in the
$\jone$ and $\jtwo$ lines from

\begin{equation}
  \tau_{u\ell} =
  \frac{1}{8\pi}\left(\frac{c}{\nu_{u\ell}}\right)^3\frac{g_u}{g_\ell}A_{u\ell}\int
  \limits_{r_{min}}^{r_{max}}[1-\exp(-E_{u\ell}/kT(r))]n_{\ell}(r)dr
  \label{eq:tau1}
  \end{equation}
with
\begin{equation}
  n_\ell(r) = \frac{1}{Z(r)}\frac{1}{\Delta \varv(r)}X_{CS}(r)n_{H_2}(r)\exp[-E_{\ell,0}/kT(r)].
  \label{eq:tau2}
  \end{equation}

In Eqs.\,\ref{eq:tau1} and \ref{eq:tau2} $n_\ell(r)$ is the radial dependent CS number
density, $Z(r)$ the partition function, $\Delta \varv(r)$ the radial dependent velocity
dispersion of the absorbing gas, $X_{CS}$ the CS abundance and $E_{\ell,0} = E_\ell -
E_0$, with $E_0$ the ground state energy. 

Applying Eqs.\,\ref{eq:tau1} and \ref{eq:tau2} to the question of the possible attenuation
of the masers requires the presence hot gas as implied by the model results. The presence
of hot gas up to $\sim$ 600 K in W51 e2 was recently demonstrated by \citet{Ginsburg2017}
from the analysis of a number of methanol lines for which $45 < E_u < 800 $ K which means
that applying Eqs.\,\ref{eq:tau1} and \ref{eq:tau2} to the case of W51 e2e is justified. 
For the temperature profile we adopt that used by \citet{Gieser2019} for the
AFGL 2591 VLA 3 hot core,
\begin{equation}
  \mathrm{T_k}(r) = 255 \,\mathrm{K}\,\left(\frac{r}{691\,\mathrm{AU}}\right)^{-0.41}
  \label{eq:tk}
\end{equation}
for $r < 4000\, \mathrm{AU} = 0.019\,\mathrm{pc}$. For $0.019\, \mathrm{pc} < r <
0.1\,\mathrm{pc}$ the kinetic temperature varies with a power law such that \tk{} = 30 K
at 0.1 pc beyond which it stays constant at 30 K.

For the \htwo{} density we assumed
\begin{equation}
  n_{\mathrm{H_2}}(r) = \left[3 \times 10^3 + n_{\mathrm{H_2}}(r_0)\left(\frac{r}{r_0}\right)^{-1}\right]\,\mathrm{cm^{-3}}
  \label{eq:nh2}
\end{equation}
where $r_0 \equiv$ 397 AU is the radial distance for the $\jtwo$ maser calculated from
Eq.\,\ref{eq:tk} for \tk{} = 320 K. This behaviour gives $\mathrm{n_{H_2}} = 3.3 \times
10^3\,\mathrm{cm^{-3}}$ at $r = 1$ pc. For the $\jone$ maser with \tk{} = 270 K, the
radial distance is found to be 601 AU and $\mathrm{n_{H_2}} \sim 10^5\,
\mathrm{cm^{-3}}$. Inspection of Fig.\,\ref{fig:beta1} shows that such a density is only
slightly larger than the maximum density for \tk{} = 270 K for which the maser brightness
temperature lies between 5000 K and 9000 K. It is rather interesting to note that the
radial separation of $\sim$ 200 AU between the masers as implied by the temperature
profile in Eq.\ref{eq:tk} agrees remarkably with observed separation of $190 \pm 60$ AU
between the masers as found by \citet{Ginsburg2019}.

The radial dependence of the velocity dispersion was related to the kinetic temperature
according to $\Delta\varv(r) = 8.0\, \mathrm{km\,s^{-1}} (\mathrm{T_k}(r)/700\,
\mathrm{K})^{0.31}$ which gives $\Delta \varv \sim 3.0\, \mathrm{km\,s^{-1}}$ at 0.1 pc
and beyond which it stayed contant at $3.0\, \mathrm{km\,s^{-1}}$.  The velocity dispersion
of 8.0 $\mathrm{km\,s^{-1}}$ at 700 K follows from the work \citet{Barr2018}. For the
radial dependence of the CS fractional abundance the result by \citet{Bruderer2009} (see
their Fig. 6) was modified to account for a higher CS fractional abundance as was argued
earlier. The fractional abundance was therefore set at $2 \times 10^{-5}$ at the position
of the $\jtwo$ maser and to decrease as a power law to $1.0 \times 10^{-7}$ at $2 \times
10^{16}$ cm. For $r > 2 \times 10^{16}$ cm the fractional abundance was set at $5 \times
10^{-9}$.

Using these radial dependencies in Eqs.\,\ref{eq:tau1} and \ref{eq:tau2} and integrating
up to $r_{max} = 1$ parsec gives $\tau = 0.6$ for the $\jone$ maser and $\tau = 1.4$ for
the $\jtwo$ maser. The difference in optical depth between the two transitions is mainly
due to the smaller radial position of the $\jtwo$ maser and therefore also the increased
density. It is also worth noting that $\sim$ 60\% of the optical depth for the $\jtwo$
maser originates inside the hot core region while for the $\jone$ maser $\sim$ 77\% of the
optical depth is due to the cold gas outside the hot core. Since the radial dependencies
used the calculation may not reflect the real situation for W51 e2e and does not consider
any clumpiness, these values should only be used as a rough estimate. In spite of the
uncertainties, these results suggest that attenuation of the $\jone$ and $\jtwo$ masers
might be non-negligible. The implication is that such attenuation may result in completely
obscuring the masers and be a contributing factor to the rarity of these masers. For W51
e2e the implication is that the detection of the masers might be due to a ``hole'' in the
surrounding molecular material as suggested by \citet{Ginsburg2019}. 

\section{Conclusions}
We solved the rate equations for the first 31 rotational states of CS in the ground
vibrational state and considered the effect of collisions, radiation and beaming on the
level populations to determine the conditions under which the $\jone$, $\jtwo$, and
$\jthree$ transitions are inverted. The main result from this study is that the three
transitions can be inverted with collisions as the only excitation mechanism. To reproduce
the brightness temperatures of the $\jone$ and $\jtwo$ masers in W51 e2e a beaming factor
of the form $(2\tau + 1)^{-2}$ is required. Brightness temperatures between 5000 K and
9000 K can be obtained for the $\jone$ maser for \htwo{} densities between
$10^5~\mathrm{cm^{-3}}$ and $10^4~\mathrm{cm^{-3}}$ and corresponding kinetic temperatures
from about 250 K to 500 K respectively. For the $\jtwo$ maser similar brightness
temperatures occur for \htwo{} densities between $10^5~\mathrm{cm^{-3}}$ and
$10^{5.5}~\mathrm{cm^{-3}}$ and corresponding kinetic temperatures between about 250 K and
500 K. A significant further result is that the two masers never have their maximum gain
at the same CS specific column density. 

No clear conclusion can be made at this time on why the CS masers are rare. It is,
however, remarkable that the CS abundance derived by \citet{Highberger2000} for the
$\vone$, $\jthree$ maser in IRC+10216 is very similar to what is inferred from the model
calculations for the $\vzero$, $\jone$ and $\jtwo$ masers. Since the CS abundance is
IRC+10216 is abnormally high, it might also be the case for W51 e2e and be the reason why
CS masers are rare. On the other hand, \citet{Barr2018} have shown that a high CS
  abundance is also present in AFGL 2591 implying that W51 e2e might not be unique in this
  regard.  It will be useful to also try to determine the CS abundance observationally in
  W51 e2e in the region where the two masers operate. We also conclude that attenuation of
  the maser emission inside and outside of the hot core might be a contributing factor for
  the non-detection of the masers.

The detection of the CS($\jone$) and CS($\jtwo$) masers in W51 e2e raises other questions
as well. For example, since the pumping of the 6.7 GHz methanol masers require the kinetic
temperature to be less than the dust temperature ($\sim$ 100 K)
\citep{Sobolev1997,Cragg2002}, what is the relation between the CS $\jone$ and the 6.7 GHz
methanol masers that are in projection rather close to each other? Furthermore, if the 6.7
GHz methanol masers indeed require maser pathlengths of about $10^{17}$ cm (6700 AU), what
physical structure are they associated with? A second question is that if a high CS
abundance is not the reason for the rarity of the CS masers, what else can be the
reason? If the rarity of the CS masers in high-mass star forming regions is indeed due to a
required high CS abundance, why is it that this requirement is met in W51 e2e and not in
the other high-mass young stellar objects in W51?

\section*{Data Availability}
No new data were generated or analysed in support of this research. The code used to
obtain the numerical results will be shared on reasonable request to the corresponding author.



\bibliographystyle{mnras}
\bibliography{maserbib1} 

\bsp	
\label{lastpage}
\end{document}